

\def\al{\alpha}
\def\eps{\epsilon}

\def\la{\lambda}

\def\si{\sigma}

\def\om{\omega}


\def\rep{representation}
\def\reps{representations}

\magnification=\magstep1 \overfullrule=0pt\parskip=6pt\baselineskip=15pt
\font\huge=cmr10 scaled \magstep2
\font\small=cmr8
 \def\la{\lambda}

\def\PS{[12]}
\def\Be{[15]}
\def\MW{[7]}
\def\MP{[20]}
\def\KP{[6]}
\def\CG{[18]}
\def\KMSW{[10]}
\def\RTW{[17]}
\def\Kac{[14]}
\def\Gan{[16]}
\def\Bou{[13]}
\def\FH{[2]}
\def\FuGO{[1]}
\def\Fuchs{[8]}
\def\GeW{[5]}
\def\Ve{[11]}
\def\WZW{[4]}
\def\Gann{[9]}
\def\RL{[19]}
\def\BMP{[3]}


{\nopagenumbers \rightline{hep-th/9412126}
\vskip2cm
\centerline{{\huge\bf Lie group weight multiplicities from conformal
field theory}}
\bigskip\bigskip\centerline{T. Gannon$^\bullet$\footnote{$^*$}{\small e-mail:
gannon@ihes.fr},  C. Jakovljevic$^\circ$ and M.A. Walton$^\circ$
{\footnote{$^\dagger$}{\small Supported in part by
NSERC. e-mail: walton@hg.uleth.ca}}}
\bigskip
\centerline{$^\bullet${\it Institut des Hautes Etudes Scientifiques}}
\centerline{{\it 91440 Bures-sur-Yvette, France}}
\bigskip
\centerline{$^\circ${\it Physics Department,
University of Lethbridge}} \centerline{{\it Lethbridge, Alberta,
Canada\ \ T1K 3M4}} \vskip2cm\centerline{{\bf Abstract}}\vskip.5cm
Dominant weight multiplicities of simple Lie groups are expressed in terms of
the modular matrices of Wess-Zumino-Witten conformal field theories, and
related objects. Symmetries of the modular matrices give rise to new relations
among multiplicities. At least for some Lie groups, these new relations are
strong enough to completely fix all multiplicities. \vfill
\noindent Short title: Weights from conformal field theory\hfill\break
\noindent PACS: 0220, 1110  \hfill\break

\eject}

\pageno=1 \noindent{{\bf 1. Introduction}}\bigskip

Wess-Zumino-Witten (WZW) models form an important class of conformal
field theories (see \FuGO, for example). They realise a current algebra
equivalent to an affine  Kac-Moody algebra. Included as a subalgebra is a
semi-simple Lie algebra, the global symmetry algebra of the theory. Not
surprisingly then, the theory of compact Lie groups and their Lie algebras has
been very useful in elucidating the properties of WZW models.

Here we present an example where the reverse is true: WZW conformal field
theories tell us something useful about Lie groups and their algebras. Weight
multiplicities are crucial numbers in the representation theory of Lie groups
(see \FH\BMP, for example). We will express the dominant weight multiplicities
of unitary highest weight \reps\ of Lie groups in terms of matrices relevant to
WZW models. We also show that our expressions give rise to new relations among
the multiplicities.

In order to explain our results, we must first discuss the connection between
modular transformations and the Lie symmetry algebra of a WZW model. Primary
fields of WZW models are in one-to-one correspondence with unitary highest
weight
\reps\ of affine Kac-Moody algebras $\hat g$ \WZW\GeW. The partition function
of a WZW
model
on a torus is a sesquilinear combination of characters of affine Kac-Moody
\reps\ \GeW. These characters transform among themselves under the action of
the modular
group of the torus \KP. Remarkably, ratios of elements of the modular $S$
matrix
equal characters of the Lie symmetry algebra $g$ evaluated at special points
(see
equation (3.5) below). This happy fact, discovered by Kac and
Peterson \KP, is the starting point of our work, and has been much exploited
elsewhere (\MW\Fuchs\Gann, for example).

Perhaps the relation between modular matrices and semi-simple Lie algebra
characters is not so surprising, if other facts are
taken into account. Consider the number of couplings between three primary
fields
of a WZW model, the so-called fusion coefficient. Since a WZW model has a Lie
symmetry
algebra, this fusion coefficient is less than or equal to the corresponding
tensor
product coefficient of the Lie algebra \GeW\KMSW. Products of Lie algebra
characters
decompose into integer linear combinations of characters, according to these
coefficients. E.\ Verlinde realised that the modular transformation matrix $S$
of a conformal field theory could tell us the fusion coefficients: products
of certain ratios of elements of the modular matrix $S$ decompose into integer
linear combinations of these same ratios, with the coefficients being the
fusion coefficients \Ve. It is reasonable that the only way such ratios could
satisfy these properties is for them to coincide with the Lie algebra
characters, but evaluated at special points.

The outline of this paper is as follows. In section 2 the work of Patera and
Sharp \PS\ is reviewed, so that it can be adapted to the use of Kac-Peterson
modular matrices in section 3. There we write a new expression for the dominant
weight multiplicities of semi-simple Lie algebras. The symmetries of the
Kac-Peterson modular $S$ matrix, and the even Weyl sums $E^{(n)}$ we introduce
in section 3, are written down in section 4. The relations between the
multiplicities
that follow are also given. Section 5 contains some simple explicit examples of
the new relations among multiplicities, and section 6 is our conclusion.

\bigskip\bigskip\noindent{{\bf 2. Lie group multiplicities from Lie
characters}}\bigskip

Define the Weyl orbit sum
$$E_\la(\si)\ :=\ {{|W(\la-\rho)|}\over{|W|}} \sum_{w\in W} \si^{w(\la-\rho)
}\ \ . \eqno(2.1)$$
Here $E$ stands for an even Weyl sum, $|W|$ is the order of the Weyl group of
$g,$ and $|W\la|$ is the order of the Weyl orbit of $\la\in P_{+}.$ The
notation is $\si^\mu={\rm exp}[-i\mu\cdot\si].$

An odd Weyl sum is the so-called discriminant
$$O_\la(\si)\ :=\ \sum_{w\in W} {det}(w)\, \si^{w\la}\ \ ,\eqno(2.2)$$
where ${det}(w)$ is the sign of the Weyl group element $w.$ The Weyl
character formula is
$$\chi_\la(\si)\ =\ O_\la(\si)/O_\rho(\si)\ \ .\eqno(2.3)$$
The character, being an even Weyl function, can be expanded in terms of the
even functions $E_\mu$ \Bou:
$$\chi_\la(\si)\ =\ \sum_{\mu\in P_{++}}\ m_\la^{\ \mu} E_\mu(\si)\ \
.\eqno(2.4)$$
The non-negative integers $m_\la^{\ \mu}$ are the dominant weight
multiplicities:
$m_\la^{\ \mu}$  denotes the multiplicity of the weight $\mu-\rho$ in the \rep\
with highest weight $\la-\rho$. Note that $m_\la^{\ \la}=1$ for all $\la\in
P_{++}$. We can consider the $m_\la^{\ \mu}$ to be the
elements of an infinite matrix $M$. If $\la-\mu\not\in {\bf Z}_{\ge}\{\al_1,
\ldots,\al_r\}$, where $\al_i$ are the simple roots, then $m_\la^{\ \mu}=0$.
Therefore $M$ is a lower triangular matrix, provided the weights are
ordered appropriately.

Not only can the Weyl character be expanded in terms of the Weyl orbit sums
$E_\la,$ but the reverse is also true:
$$E_\la(\si)\ =\ \sum_{\mu\in P_{++}} \ell_\la^{\ \mu} \chi_\mu(\si)\ \ .
\eqno(2.5)$$
The coefficients $\ell_\la^{\ \mu}$ are easily shown to be integers, but are
{\it not} in general non-negative. However
if $L$ is the matrix with elements $\ell_\la^{\ \mu},$ then clearly $M=L^{-1}$.
So, if the triangular matrix $L$ can be calculated, it is a simple matter to
invert it to obtain the dominant weight multiplicities \PS.

Patera and Sharp \PS\ also point out that the equations above allow the
calculation of the $\ell_\la^{\ \mu}$ for fixed $\la$ using the Weyl group.
This
corresponds to the following formula,
$$\ell_\la^{\ \mu} = {{|W(\la-\rho)|}\over{|W|}}\ \sum_{w,x\in W} {det}(w)\,
\delta_{w\rho+x(\la-\rho)}^\mu\ \ , \eqno(2.6)$$
that can be derived from the defining relation for the $\ell.$

\bigskip\bigskip\noindent{{\bf 3. Lie group multiplicities from WZW modular
matrices}}\bigskip

Define
$$O^{(n)}_\la(\si)\ :=\ F_n \sum_{w\in W} {det}(w) \,\si_{(n)}^{w\la}\ =\
S^{(n)}_{\la\si}\ \ ,\eqno(3.1)$$
and
$$E_\la^{(n)}(\si)\ :=\ {{|W(\la-\rho)|}\over{|W|}}\sum_{w\in W}
\si_{(n)}^{w(\la-\rho)}\ \ ,\eqno(3.2)$$
with
$$\si_{(n)}^{\mu}\ :=\ {\rm exp}[-2\pi i\mu\cdot\si/n]\ ,\ \ \ \
F_n\ :=\ {i^{|\Delta_+|}\over n^{r/2}\sqrt{|M^*/M|}}\ \ ;\eqno(3.3)$$
where $|\Delta_+|$ is the number of positive roots of $g$, and $M$ here is
the weight lattice.
The matrix $S^{(n)}$ in eq.(3.1) is the Kac-Peterson modular matrix
of WZW models, corresponding to the affine algebra $\hat{g}$ at
level $k=n-h^\vee$, where $h^\vee$ is the dual Coxeter number of $g$. Define
$$P_{+}^n\ :=\ \{\sum_{i=0}^n\la_i\om^i\,|\,\la_i\in{\bf Z}_{\ge}\ ,\
\sum_{i=0}^ra_i^\vee\la_i=n\,\}\ ,\eqno(3.4)$$
and $P_{++}^n$ similarly, except with ${\bf Z}_{\ge}$ replaced with
${\bf Z}_>$. The positive integers $a_i^\vee$ in eq.(3.4) are the {\it
co-marks} \Kac\ of $\hat g$.
We read from eq.(3.1)
that
$$S^{(n)}_{\la\si}/S^{(n)}_{\rho\si}\ =\ O^{(n)}_\la(\si)/O^{(n)}_\rho(\si)\
=:\ \chi^{(n)}_\la(\si)\ ,\eqno(3.5)$$
where
$$\chi^{(n)}_\la(\si)\ =\ \chi_\la(2\pi\,\si/n)\ .\eqno(3.6)$$
This is the ``happy fact'' referred to in the introduction.
 $M$ and $L$ are lower triangular; therefore whenever $\la\in P_{++}^n$,
this  remarkable relation implies
$$\chi^{(n)}_\la(\si)\ =\ \sum_{\mu\in P_{++}^n} m_\la^{\ \mu}E^{(n)}_\mu(\si)\
\ ,\eqno(3.7)$$
and
$$E^{(n)}_\la(\si)\ =\ \sum_{\mu\in P_{++}^n}
\ell_\la^{\ \mu}\chi^{(n)}_\mu(\si)\ \ .\eqno(3.8)$$
Eq.(3.8) also holds for $\la\in P_{++}\cap P_+^n$.
Using the unitarity of the modular $S$ matrix
$$\sum_{\nu\in P_{++}^n} O^{(n)}_\la(\nu)\,O^{(n)*}_\mu(\nu)\ =\
\sum_{\nu\in P_{++}^n} S^{(n)}_{\la\nu} \,S^{(n)*}_{\nu\mu}\ =\
\delta_\la^\mu\ ,\eqno(3.9)$$
we arrive at
$$\ell_\la^{\ \mu}\ =\ \sum_{\si\in P_{++}^n}
E^{(n)}_\la(\si)\,O^{(n)}_\rho(\si)
\,O^{(n)*}_\mu(\si)\ =\ \sum_{\si\in P_{++}^n} E^{(n)}_\la(\si)\,
S^{(n)}_{\rho\si} \,S^{(n)*}_{\mu\si}\ \ ,\eqno(3.10)$$
valid whenever both $\la\in P_{++}\cap P_+^n$ and $\mu\in P_{++}^n$.
Eq.(3.10) can be generalized:
$$\sum_{\si\in P_{++}^n}\ell_\la^{\ \si}N_{\mu\si}^{(n)\nu}=\sum_{\si\in
P_{++}^n}E_\la^{(n)}(\si)\,S^{(n)}_{\mu\si}\,S^{(n)*}_{\nu\si}\ ,\eqno(3.11)$$
where $N_{\mu\si}^{(n)\nu}$ are the WZW fusion rules, which we may take to
be defined by Verlinde's formula \Ve :
$$N^{(n)\ \ \nu}_{\la\mu}\ =\ \sum_{\si\in P^{n}_{++}}\ \chi_\la^{(n)}(\si)
\,S^{(n)}_{\mu\si} \,S^{(n)*}_{\nu\si}\ \ \ .\eqno(3.12)$$
Of course we also get directly from
eq.(3.7) that
$$N_{\la\mu}^{(n)\nu}=\sum_{\si,\gamma\in P_{++}^n}m_\la^{\ \gamma}\,
E_\gamma^{(n)}(\si)\,S_{\mu\si}^{(n)}\,S_{\nu\si}^{(n)*}\ .\eqno(3.13)$$

Because $L$ is lower triangular, an easy argument gives
$$m_\la^{\ \mu}=(L^{(n)-1})_\la^{\ \mu}\eqno(3.14)$$
for all $\la,\mu\in P_{++}^n$, where $L^{(n)}$ is defined to be the
sublattice of $L$ obtained by restricting it to the set $P_{++}^n$.
Thus equations like eq.(3.10) provide a simple method of calculating dominant
weight multiplicities $m_\la^{\ \mu}.$ Moreover, if we find a permutation $\pi$
of $P_{++}^n$ which commutes with both $S^{(n)}$ and $E^{(n)}$, then it will
be an exact symmetry of both $\ell$ and $m$. More generally, if $S^{(n)}$
and $E^{(n)}$ both transform ``nicely'' under a permutation $\pi$ of
$P_{++}^n$, then we can expect to derive new relations for $\ell$ and $m$.
This is the motivation for the following section.

\bigskip\bigskip\noindent{{\bf 4. New relations between
multiplicities}}\bigskip

In this section, we will show that symmetries of the Kac-Peterson modular
matrices $S^{(n)}$ of WZW models give rise to new relations between
finite-dimensional Lie algebra multiplicities.

The most obvious symmetry concerns the affine Weyl group $\widehat{W}$ of
$\hat{g}$. We know \Kac\ that it is isomorphic to the semi-direct product of
the
(finite) Weyl group $W$ with the group of translations in the coroot lattice
$Q^\vee$. We also know that the $\widehat{W}$-orbit of any weight
intersects $P_{+}^n$ in exactly one point. More precisely,
let $\la\in M$ be some weight. Then there exists an element $\al$ in the
coroot lattice of $g$, and some $w\in W$, such that
$$[\la]\ :=\ w(\la+n\al)\ \in\ P_{+}^n\ \ .\eqno(4.1)$$
We will use this observation throughout this section. $[\la]$ is uniquely
determined by $\la$ (and $n$), but $w$ will be only if $[\la]\in P_{++}^n$.
Define $\eps
(\la):=0$ if $[\la]\not\in P_{++}^n$, and $\eps(\la):=det(w)$ otherwise, where
$w\in W$ satisfies eq.(4.1).

We read directly from eqs.(3.1),(3.2) respectively that
$$\eqalignno{S_{\la\mu}^{(n)}=&\ \eps(\la)\,S_{[\la]\mu}^{(n)}=\eps(\mu)\,
S_{\la[\mu]}^{(n)}\ ;&(4.2)\cr
\cr
E_{\la}^{(n)}(\si)=&\ E_{\la}^{(n)}([\si])= {|W(\la-\rho)|\over |W([\la-
\rho])|}\,E_{[\la-\rho]+\rho}^{(n)}(\si)\ .&(4.3)\cr}$$
By the argument which gave us eq.(3.10), we find that for any $\la\in P_{++}$,
$\mu\in P_{++}^n$,
$$\sum_{{\nu\in P_{++}\atop [\nu]=\mu}}\eps(\nu)\,\ell_\la^{\ \nu}=\sum_{\si
\in P_{++}^n}E^{(n)}_{\la}(\si)\,S^{(n)}_{\rho\si}\,S^{(n)*}_{\mu\si}\ .
\eqno(4.4)$$
Thus for any $\mu\in P_{++}^n$, and any  $\la\in P_{++}$ with $[\la-\rho]+
\rho\in P_{+}^n$, we get the truncation
$$\ell_{[\la-\rho]+\rho}^{\
\mu}={|W([\la-\rho])|\over|W(\la-\rho)|}\sum_{{\nu\in
P_{++}\atop [\nu]=\mu}}\eps(\nu)\,\ell_\la^{\ \nu}\ . \eqno(4.5)$$
Roughly speaking, eq.(4.5) says that if we know the $\ell$'s for ``large''
weights, then we know them for ``small'' ones. Incidently, if $[\la-\rho]+\rho
\not\in P_+^n$, then (4.5) holds if we replace its LHS with a sum similar
to that of its RHS. Similar comments hold below if $\pi_A(\la)\not\in P_+^n$
in (4.13), or $\pi_a(\la)\not\in P_+^n$ in (4.17).

Next, consider the symmetries involving the outer automorphisms of affine Lie
algebras $\hat g,$ or equivalently, the automorphisms of the {\it extended}
Coxeter-Dynkin diagrams of $g.$ If an outer automorphism is also a symmetry of
the unextended Coxeter-Dynkin diagram of $g$, i.e.\ it fixes the extended
node, then it is well known to be an exact symmetry $C$ (a {\it conjugation})
of both the $\ell$'s and $m$'s:
$$m_{C\la}^{\ C\mu}=m_\la^{\ \mu}\ , \qquad \ell_{C\la}^{\ C\mu}=\ell_\la^{\
\mu}\ .
\eqno(4.6)$$
We are interested here instead in those automorphisms which are not
 conjugations. Denote such an automorphism by $A,$
and the fundamental weights of $\hat g$ by $\om^i,$ with $i=0,1,2,\ldots,r.$
There is one of these automorphisms $A=A_i$ for every node of the extended
diagram with mark $a_i=1$; $A_i$ will send $\om^0$ to $\om^i$. Since
$$A(\la-n\om^0)\ =\ w_A(\la-n\om^0)\ ,\eqno(4.7)$$
with $w_A$ an element of the Weyl
group $W$ of $g,$ for all $\la\in P_{+}^n,$ we have \Be
$$S^{(n)}_{A\la,\si}\ =\ S^{(n)}_{\la\si}\ {\rm exp}\left[-2\pi i\,
(A\om^0)\cdot \si\right]\,det(w_A)=S^{(n)}_{\la\si}\,\exp[-2\pi i\,
(A\om^0)\cdot(\si-\rho)]\ \ .\eqno(4.8)$$
Similarly
$$\eqalignno{E^{(n)}_\la(A\si)\ =&\ E^{(n)}_\la(\si)\,
\exp[-2\pi i\,(A\om^0)\cdot (\la-\rho)]\ ,&(4.9)\cr
\cr E^{(n)}_{A\la}(\si)\  =&\ E^{(n)}_{\pi_A\la}(\si)\ {{|W(A\la-\rho)|}\over
{|W(\pi_A(\la)-\rho)|}}\  {\rm exp} \left[-2\pi i\,(A\om^0)\cdot
\si\right]\ \ ,&(4.10)\cr}$$  where $\pi_A$ denotes the one-to-one map from
$P_{++}^n$ to $P_{++}$ given by
$$\pi_A(\la):=[\la-w_A^{-1}\rho]+\rho\ .\eqno(4.11)$$
For fixed $g$, $w_A$ and hence the map $\pi_A$ is readily obtained from
eq.(4.7) -- $\pi_A$ will be the identity only when $A$ is. For example, for
$g=su(r+1)$ and $A=A_j$ satisfying $A\om^{i}=\om^{i+j}$, we get
$$\bigl(w_A\la\bigr)_i\ =\ \left\{\matrix{\la_{i-j}&{\rm if}\ i\ne j\cr
-\la_1-\cdots-\la_r&{\rm if}\ i=j\cr}\right. \ ,\eqno(4.12)$$
$det(w_A)=(-1)^{rj}$, and $\pi_A(\la)=\la+(r+1)\,\om^{r+1-j}$.

{}From our main result (3.10), we immediately find
$$\ell_{A\la}^{\ A\mu}\ =\ \ell_{\pi_A(\la)}^{\ \mu}\
{{|W(A\la-\rho)|}\over{|W(\pi_A
\la-\rho)|}}\ det(w_A)\ ,\eqno(4.13)$$
for any $\la,\mu\in P_{++}^n$, provided $\pi_A(\la)\in P_+^n$. Unfortunately
 $\pi_A$ will only be a permutation of $P_{++}^n$ in the trivial case when
$A=id.$, so it is not easy to see what eq.(4.13) directly implies for the
$m_\la^{\ \mu}$.

There are also Galois symmetries of the Kac-Peterson modular matrix $S^{(n)},$
first discovered in \Gan \RTW\ (and generalized to all rational conformal field
theories in \CG). The $S^{(n)}_{\la\mu}/F_n$ and $E^{(n)}_\la(\si)$ are
polynomials with rational coefficients in a primitive $(nN)$-th root of unity,
where $N=|M^*/M|^{{1\over2}},$ $M$ here being the weight  lattice of $g.$ So,
any polynomial relation involving them and rational numbers only, will also be
satisfied if this primitive $(nN)$-th root of unity is replaced by another.

Let $a$ be an integer coprime to $nN,$ and let $a(S^{(n)})$ denote the
Kac-Peterson matrix after the primitive $(nN)$-th root of unity is replaced by
its $a$-th power (ignoring here the irrelevant factor $F_n$). For such
$a,$ and for $\la\in P_{++}^n,$ recall the quantities
$[a\la]\in P_{++}^n$ and $\eps(a\la)\in\{\pm 1\}$ defined around eq.(4.1).
For each $a$ coprime to
$nN$, the map $\la\mapsto [a\la]$ is a permutation of $P_{++}^n$. From the
form of the  matrix $S^{(n)},$ it is easy to find
$$a\left(S^{(n)}_{\la\mu}\right)\ =\ \eps(a\la)\,
S^{(n)}_{[a\la],\mu}\ =\ \eps(a\mu) \,S^{(n)}_{\la,[a\mu]}\ \ .\eqno(4.14)$$
In a similar fashion, we find
$$a\left(E^{(n)}_\la(\si)\right)\ =\ {|W(\la-\rho)|\over |W(\pi_a\la-\rho)|}\
E^{(n)}_{\pi_a(\la)}(\si)\ =\ E^{(n)}_\la([a\si])\ \ ,\eqno(4.15)$$
where $\pi_a$ denotes the one-to-one map from $P_{++}^n$ to $P_{++}$
defined by
$$\pi_a(\la):=[a\la-a\rho]+\rho\ .\eqno(4.16)$$

A little work yields
$$\ell_\la^{\ \mu}\ =\ \eps(a\mu)\,\eps(a\rho)\ {|W(\la-\rho)|\over
|W(\pi_a\la-
\rho)|}\ \sum_{\nu\in P_{++}^n}\
\ell_{\pi_a(\la)}^{\ \nu} \,N^{(n)\ \ \ [a\mu]}_{\nu,[a\rho]}\ \ ,\eqno(4.17)$$
whenever $\mu\in P_{++}^n$ and $\la,\pi_a(\la)\in P_{++}\cap P_{+}^n$. Here we
have used the Verlinde formula \Ve\ (3.12)
for the fusion coefficients $N^{(n)\ \nu}_{\la\mu}.$ Let
$N^{(n)}_\la$ denote the {\it fusion matrix} defined by $(N^{(n)}_\la
)_\mu^\nu:=N^{(n)\ \nu}_{\la\mu}$. The matrix $N^{(n)}_{[a\rho]}$ will always
be invertible \CG, so eq.(4.17) tells us that for any fixed $\la\in P_{++}$,
the values $\ell_{\pi_a(\la)}^{\ -}$ will be known once the $\ell_\la^{\ -}$
are
known, and conversely, provided $a$ and $n$ satisfy the usual conditions.
Eq.(4.17) can also be interpreted as an expression for the fusion matrices
$N^{(n)}_{[a\rho]}$ in terms of the $\ell$'s and $m$'s.

It is again difficult to express this Galois symmetry directly at the level
of the multiplicities $m_\la^{\ \mu}$ themselves. But if
$\pi_a$ {\it is} a permutation of $P_{++}^n$, we get from eq.(4.17):
$$m_\la^{\ \mu}=\ \eps(a\mu)\,\eps(a\rho)\ {|W(\pi_a\la-\rho)|\over |W(\la-
\rho)|}\,\sum_{\nu\in P_{++}^n}\ (N^{(n)-1
}_{[a\rho]})_{\ [a\la]}^\nu\,m_\nu^{\ \pi_a(\mu)}\ .\eqno(4.18)$$

A special case of eq.(4.17) occurs when $[a\rho]=\rho.$ (More generally, a
similar simplification happens whenever $[a\rho]=A\rho$ for some outer
automorphism $A.$) Then $\eps(a\rho)=\eps(a\la)$ for all $\la\in P_{++}^n$
(apply eq.(4.14) with $\mu=\rho$, together
with the fact that $S_{\rho\nu}^{(n)}>0$ for all $\nu\in P_{++}^n$). Eq.(4.17)
reduces to
$$\ell_\la^{\ \mu}={|W(\la-\rho)|\over |W(\pi_a\la-\rho)|}\,\ell_{\pi_a(\la)}^
{\ [a\mu]}\ .\eqno(4.19)$$
For example this happens whenever $a=-1$, and we get an example of eq.(4.6).

Similarly, suppose $[a\mu]=A\rho$, for some outer automorphism $A$. Then
provided $\la,\pi_a(\la)\in P_{++}\cap P_+^n$, eq.(4.17) reduces to
$$\ell_\la^{\ \mu}=\eps(a\mu)\,\eps(a\rho)\,{|W(\la-\rho)|\over |W(\pi_a\la-
\rho)|}\,\ell_{\pi_a\la}^{\ A[a\rho]}\ .\eqno(4.20)$$

Another noteworthy special case of eq.(4.17) involves those weights $\la'$ with
the property that $\ell_{\la'}^{\ \mu}=\delta_{\la'}^\mu$ for all $\mu\in
P_{++},$
i.e. those $\la'$ for which $\la'-\rho$ is a miniscule weight. For
$su(r+1)$, they are the fundamental weights $\la'=\om^i+\rho$. Then for any
$\la\in P_{++}^{(n)}$ with $\pi_a(\la)=\la',$
$$\ell_\la^{\ \mu}=\eps(a\mu)\,\eps(a\rho)\,{|W(\la-\rho)|\over |W(\la'-
\rho)|}\,N^{(n)\ \ [a\mu]}_{\la',[a\rho]}\ ,\eqno(4.21)$$
if, as usual,
$a$ is coprime to $nN$ and $\la',\mu\in P_{++}^n$.
The fusions involving the fundamental weights $\om^i+\rho$ are easy to
compute, so the RHS of eq.(4.21) can be explicitly evaluated in all cases.
One of the reasons equations (4.17)-(4.21) could be interesting
is because they suggest the rank-level duality that
WZW fusions satisfy \RL\ could appear in some way in the $\ell$'s
and $m$'s.

\eject\noindent{{\bf 5. Examples: the case of $g=su(3)$}}\bigskip

For concrete illustrations of our results, we will focus
on the example of $g=su(3)$. The comments in this section can be
extended to any $su(r+1)$ without difficulty.

Consider first eq.(4.13). It reduces to
$$\ell_{(\la_1,\la_2)}^{\ (\mu_1,\mu_2)}=\ell_{(n-\la_1-\la_2+3,\la_2)}^{\
(n-\mu_1
-\mu_2,\mu_2)}\ ,\qquad\forall n\in {\bf Z}_>\ , \eqno(5.1)$$
provided only that: $\mu_1+\mu_2<n$;\ $\la_2,\mu_1,\mu_2\ge 1$; and either
$\la_1+\la_2<n$, $\la_1\ge 3$, or $\la_1+\la_2\le n$, $\la_1>3$.

Eq.(5.1) is very powerful. For example, suppose $\la_1\ge 3$. Put $n=\la_1+
\la_2+1$. Then eq.(5.1) reduces $\ell_\la^{\ \mu}$ to $\ell_{\la'}^{\ \mu'}$,
where
$\la'=(4,\la_2)$. In fact, if both $\la_1,\la_2\ge 3$ and $\mu\in P_{++}$,
we obtain from eq.(5.1) (once we compute the values of $\ell_{(4,4)}^{\ -}$):
$$\ell_\la^{\ \mu}=\left\{\matrix{+1&{\rm if}\ \mu\in\{\la,\ (\la_1-3,\la_2),\
(\la_1,\la_2-3)\}\cr
-1&{\rm if}\ \mu\in\{(\la_1-2,\la_2+1),\ (\la_1+1,\la_2-2),\ (\la_1-2,\la_2
-2)\}\cr
0&{\rm otherwise}\cr}\right. \ .\eqno(5.2)$$
In fact, using eq.(4.5),
we find that eq.(5.2) gives the correct value of $\ell_\la^{\ \mu}$ for
any $\la,\mu\in P_{++}$ (whether or not $\la_1\ge3$), with one exception:
$\ell_{(2,2)}^{\ (1,1)}=-2\ne 0$,

Next, let us turn to the Galois symmetries. It is not difficult to see that
$[a\rho]=\rho$ only for $a\equiv\pm 1$ (mod $n$), in which case $\pi_a(\la)
=\pi_A(\la)$ for some outer automorphism $A$. Thus for $g=su(3),$ eq.(4.19)
only gives information also obtainable from eq.(5.1).

However, eqs.(4.17),(4.20),(4.21) are very
strong here. To give one explicit
example, consider $\la=(4,4)$, $a=5$, $n=8$. Then $\pi_a\la=(2,2)$. The
only possible nonzero elements of $\ell_{(2,2)}^{\ -}$ are
$\ell_{(2,2)}^{\ (2,2)}=
1$ and $\ell_{(2,2)}^{\ (1,1)}$. We get from eq.(4.17) that e.g.\
$$\eqalignno{\ell_{(4,4)}^{\ (3,3)}&\ =0\ ,\quad \ell_{(4,4)}^{\ (1,1)}=N^{(8)\
\
(3,3)}_{(2,2),\,(3,3)}+\ell_{(2,2)}^{\ (1,1)}\ ,&(5.3)\cr
\ell_{(4,4)}^{\ (2,2)}&=-N^{(8)\ \ (2,2)}_{(2,2),\,(3,3)}\ ,\quad
\ell_{(4,4)}^{\
(5,2)}=-N^{(8)\ \ (2,5)}_{(2,2),\,(3,3)}\ ,\quad \ell_{(4,4)}^{\ (1,4)}=
N^{(8)\ \ (1,4)}_{(2,2),\,(3,3)}\ .&\cr}$$

Note that the relations (4.5), (4.13), and (4.17),
together with the selection rule ``$\ell_\la^{\ \mu}\ne 0$ requires $\la-\mu\in
{\bf Z}_{\ge}\{\al_1,\ldots,\al_r\}$'', and the normalization $\ell_\la^{\
\la}=
1$, easily determine all values of $\ell_\la^{\ \mu}$ and
hence $m_\la^{\ \mu}$. In particular, we have seen that eqs.(4.5) and (4.13)
determine all $\ell_\la^{\ \mu}$ provided the values $\ell_{(4,4)}^{\ -}$ are
known.
Eq.(5.3) fixes all values of $\ell_{(4,4)}^{\ -}$ except $\ell_{(4,4)}^{\
(1,1)}$,
but by eq.(5.1) we find $\ell_{(4,4)}^{\ (1,1)}=\ell_{(4,4)}^{\ (7,1)}=0$. A
similar conclusion should apply to any $su(r+1)$.

\eject\noindent{{\bf 6. Conclusion}}\bigskip

Our main result is the expression (3.10) for the inverse of the matrix of
dominant
weight multiplicities, in terms of the Kac-Peterson modular matrix $S^{(n)}$
and
the even Weyl orbit sums $E_\la^{(n)}(\si).$ Surely the even Weyl orbit sums
will find other uses in the study of WZW models and affine Kac-Moody algebras.

Also obtained were relations (4.5), (4.13) and (4.17)
among the ``inverse multiplicities'' $\ell_\la^{\ \mu},$ that are consequences
of
the symmetries of the Kac-Peterson matrices and the $E_\la^{(n)}(\si).$
Relations between the dominant weight multiplicities $m_\la^{\ \mu}$ follow in
certain cases, as eq. (4.18) shows. These relations would be difficult to
understand from the point of view of Lie groups and their semi-simple Lie
algebras only, but are quite natural in WZW models, or in their affine
Kac-Moody current algebras. The new relations
arise because the semi-simple symmetry algebra of a WZW
model is a special subalgebra of its affine current algebra. The affine Weyl
and outer
automorphism symmetries are especially natural in  the Kac-Moody context.
Perhaps the Galois symmetries can be better understood in terms of affine
Kac-Moody algebras.

The debt we owe to the previous work of Patera and Sharp \PS\ is obvious when
comparing sections 2 and 3. We should also mention a paper by Moody and Patera
\MP. In it class functions of an arbitrary semi-simple compact Lie group are
decomposed into sums of irreducible characters. The method uses the characters
of elements of finite order (EFO's) of the group to approximate the characters
of arbitrary elements. These characters of EFO's have a
marked similarity to the Kac-Peterson ratios $\chi^{(n)}_\la(\si).$ Moody and
Patera also use modular arithmetic to ensure that the approximation leads to
the correct answers, if a suitable set of EFO's is chosen. In a similar way, we
automatically recover the exact weight multiplicities from the Kac-Peterson
ratios, which are the Lie algebra characters evaluated at special
points.

Further comparison with \MP\ is clearly warranted. Perhaps it will help us
toward a better understanding of WZW models, one that approaches the
current understanding of semi-simple compact Lie groups and their algebras.

\bigskip\bigskip\noindent{{\bf Acknowledgements}}\bigskip

T.G. and M.W. thank the IHES for hospitality, and M.W. also thanks R.T. Sharp
and P. Mathieu for discussions.

\eject\noindent{{\bf References}}\bigskip

\item{1.} Goddard P and Olive D\ 1986\ {\sl Int. J. Mod. Phys.}\
{\bf A1}\ 303\hfill\break
Fuchs J\ 1992 {\sl Affine Lie Algebras and Quantum Groups} (Cambridge:
Cambridge University Press)

\item{2.} Fulton W and Harris J\ 1991 {\sl Representation Theory - A First
Course} (New York: Springer-Verlag)

\item{3.} Bremner M\ R, Moody R\ V and Patera J\ 1985 {\sl Tables of Dominant
weight multiplicities for representations of simple Lie algebras} (New York:
Marcel Dekker)

\item{4.} Witten E\ 1984
{\sl Nucl. Phys.}\ {\bf 92} 455\hfill\break
Knizhnik V and Zamolodchikov A\ B\ 1984 {\sl Nucl. Phys.}\ {\bf B247} 83

\item{5.} Gepner D and Witten E\ 1986 {\sl Nucl. Phys.}\ {\bf B278} 493

\item{6.} Kac V\ G and Peterson D\ 1984 {\sl Adv. Math.} {\bf 53} 125

\item{7.} Walton M\ A\ 1990 {\sl Phys. Lett.} {\bf 241B} 365;\ {\sl
Nucl. Phys.} {\bf B340} 777

\item{8.} Fuchs J\ 1991 {\sl Comm. Math. Phys.}\ {\bf 136} 345

\item{9.} Gannon T\ 1994 The classification of $su(m)_k$ automorphism
invariants, IHES preprint (hep-th/9408119)

\item{10.} Kirillov A\ N, Mathieu P, S\'en\'echal D and Walton M\
A\ 1993 {\sl Nucl. Phys.}\ {\bf B391} 257

\item{11.} Verlinde E\ 1988 {\sl Nucl. Phys.}\ {\bf B300} 360

\item{12.} Patera J and Sharp R\ T\ 1989 {\sl J.\ Phys.}\ {\bf A22}
2329

\item{13.} Bourbaki N\ 1968 {\sl Groupes et alg\`ebres de Lie} Chapitres IV-VI
(Paris: Hermann)

\item{14.} Kac V\ G\ 1990 {\sl Infinite dimensional Lie algebras} 3rd
ed. (Cambridge: Cambridge University Press)

\item{15.} Bernard D\ 1987 {\sl Nucl. Phys.} {\bf B288} 626

\item{16.} Gannon T\ 1993 {\sl Nucl. Phys.}\ {\bf B396} 708

\item{17.} Ruelle Ph, Thiran E and Weyers J\ 1993 {\sl Nucl. Phys.} {\bf
B402} 693

\item{18.} Coste A and Gannon T\ 1994 {\sl Phys. Lett.} {\bf 323B} 316

\eject\item{19.} Goodman F and Wenzl H\ 1990 {\sl Adv. Math.}\ {\bf 82}\
244 (1990)\hfill\break Cummins C\ J\ 1991 {\sl J. Phys.}\ {\bf A24} 391
\hfill\break
Mlawer E, Naculich S, Riggs H and Schnitzer H\ 1991 {\sl Nucl. Phys.}\
{\bf B352} 863

\item{20.} Moody R\ V and Patera J\ 1987 {\sl Mathematics of Computation}
{\bf 48} 799

\bye